\documentclass[9pt,twocolumn,twoside]{osajnl}
\usepackage[ justification=justified]{caption} 
\usepackage{multirow}
\usepackage{titlesec}
\graphicspath{{images/}{../images/}}

\journal{optica} 
\setboolean{shortarticle}{false}
\usepackage{lineno}
\title{ Long-distance continuous-variable quantum key distribution over 100 km fiber with local local oscillator}

\author[1,*]{Adnan A.E. Hajomer}
\author[1,2]{Ivan Derkach}
\author[1]{Nitin Jain}
\author[1,3]{Hou-Man Chin}
\author[1]{Ulrik L. Andersen}
\author[1,$\dagger$]{Tobias Gehring}

\affil[1]{Center for Macroscopic Quantum States (bigQ), Department of Physics, Technical University of Denmark, 2800 Kongens Lyngby, Denmark}
\affil[2]{Department of Optics, Faculty of Science, Palacky University, 17. listopadu 12, 771 46 Olomouc, Czech Republic}
\affil[3]{Department of Photonics, Technical University of Denmark, 2800 Kongens Lyngby, Denmark}
\affil[*]{Corresponding authors: * aaeha@dtu.dk, $^\dagger$ tobias.gehring@fysik.dtu.dk}

\begin{abstract}
Quantum key distribution (QKD) enables two remote parties to share encryption keys with security based on the laws of physics. Continuous variable (CV) QKD with coherent states and coherent detection integrates well with existing telecommunication networks. However, thus far, long-distance CV-QKD has only been demonstrated using a highly complex scheme where the local oscillator is transmitted, opening security loopholes for eavesdroppers and limiting its potential applications. Here, we report a long-distance CV-QKD experiment with a locally generated local oscillator over a 100 km fiber channel with a total loss of 15.4 dB. This record-breaking distance is achieved by controlling the phase-noise-induced excess noise through a machine-learning framework for carrier recovery and optimizing the modulation variance. We implement the full CV-QKD protocol and demonstrate the generation of keys secure against collective attacks in the finite-size regime. Our results mark a significant milestone for realizing CV quantum access networks with a high loss budget, and pave the way for large-scale deployment of secure QKD. 
\end{abstract}

\setboolean{displaycopyright}{false}

\begin{document}

\maketitle
\section{Introduction}

\begin{figure*}[h]
\centering
\includegraphics[width=\linewidth]{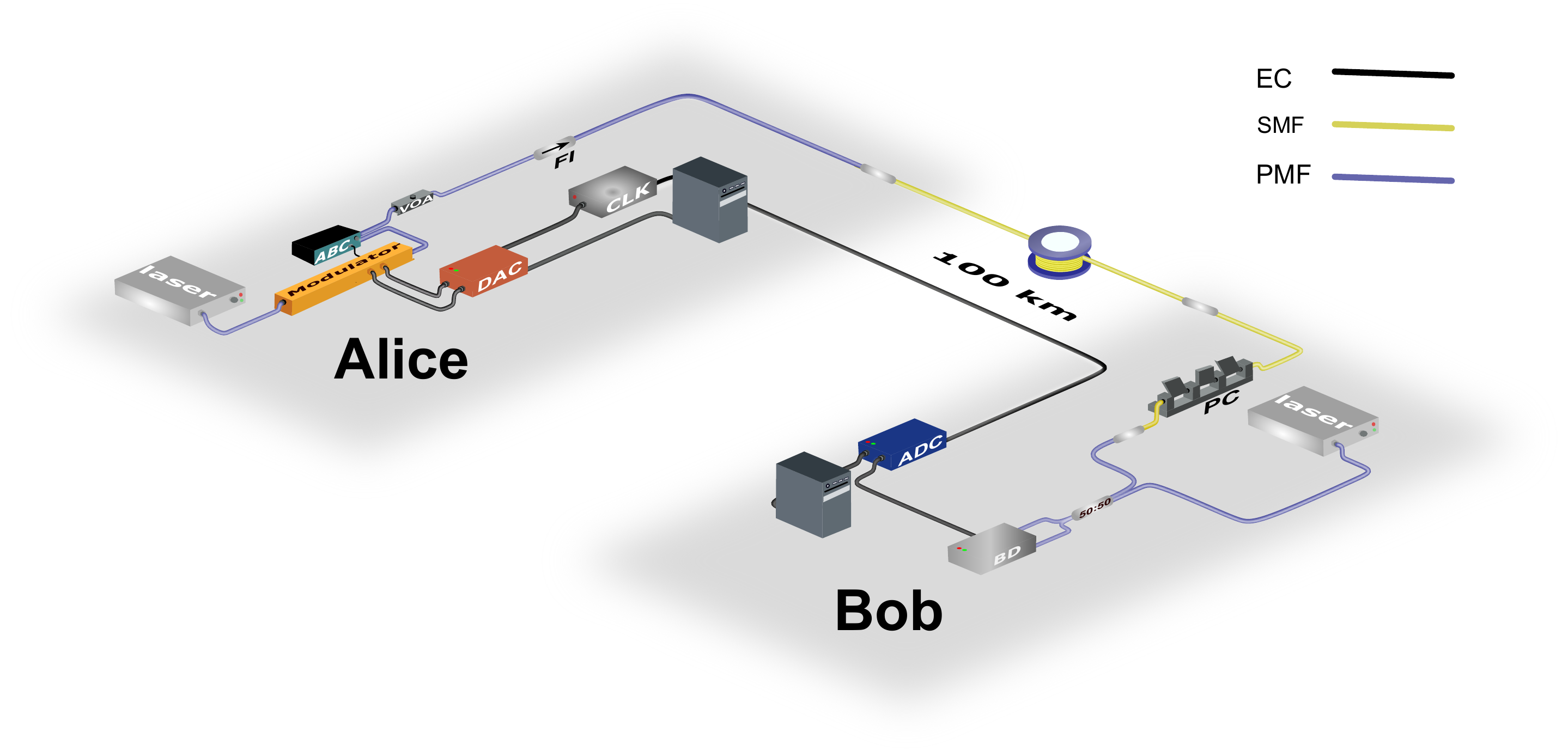}
\caption{\textbf{Long distance CV-QKD system}. Alice's station consists of a continuous wave (CW) laser operating at 1550 nm, an in-phase and quadrature (IQ) modulator with an automatic bias controller (ABC) for producing coherent states at sideband frequencies. A digital-to-analog converter (DAC) with a resolution of 16 bits and a sampling rate of 1 Gsample/s was used to drive the IQ modulator. A variable optical attenuator (VOA) was used after the IQ modulator to adjust the modulation variance of the quantum signal. A Faraday isolator (FI), whose forward direction is indicated by the arrow, is used before a 100 km ultra-low-loss fiber channel that constitutes the quantum channel. Bob's station consists of a polarization controller (PC) to adjust the polarization of the incoming signal and a balanced beamsplitter to overlap this signal with a local oscillator generated from another CW laser (unlocked/free-running with respect to Alice's laser). The signal was detected and digitized using a balanced detector (BD), followed by an analog-to-digital converter (ADC) with a sampling rate of 1 Gsample/s.} 
\label{fig:1}
\end{figure*}

Secure exchange of cryptographic keys over public channels is a critical prerequisite for maintaining secure communication. Currently, this is often accomplished using public key cryptography based on computationally hard problems such as integer factorization and (elliptic curve) discrete logarithm, providing computational security~\cite{hellman1976new,rivest1978method}. However, the emergence of advanced algorithms and quantum computers threatens the security of these methods~\cite{shor1994algorithms, arute2019quantum}. Quantum key distribution (QKD) offers a promising solution using the principles of quantum physics to share information-theoretically secure keys between remote users~\cite{BEN84}. However, the transmission range of QKD remains limited due to the inverse scaling of the secret key rate with transmission distance~\cite{pirandola2017fundamental}, necessitating the use of trusted or untrusted nodes. Extending the distance between these nodes is vital for large-scale deployment of QKD.

While there have been numerous lab demonstrations and field trials for point-to-point long-distance QKD using discrete variable (DV) protocols~\cite{pirandola2020advances}, continuous variables (CV) encoding of quantum information, such as the amplitude and the phase quadrature of the electromagnetic field of light, offers a powerful approach for secure communication~\cite{grosshans2002continuous, weedbrook2004quantum}. This is because CV-QKD systems can be constructed using components found in coherent optical telecommunication systems, including in-phase and quadrature modulators for quantum state preparation and coherent detection facilitated by a local oscillator (LO) for quantum state measurement. However, two major challenges in CV-QKD limit the transmission distance: excess noise~\cite{lodewyck2005controlling}, mainly originating from the laser's phase noise and limited classical information reconciliation efficiency~\cite{leverrier2008multidimensional}.

To control the excess noise due to the laser phase noise, long-distance CV-QKD demonstrations usually transmit the local oscillator (TLO) from the transmitter to the receiver~\cite{lodewyck2007quantum, jouguet2013experimental, wang201525, huang2016long, zhang2020long}. In such an implementation, the quantum state and the LO are prepared from the same laser source and propagate through an insecure quantum channel, ensuring a stable relative phase between the LO and the quantum signal. However, this configuration exposes the LO to potential adversaries, enabling side-channel attacks~\cite{ma2013local,jouguet2013preventing}, and necessitates complex multiplexing techniques to avoid cross-talk from the strong TLO signal to the fragile quantum states~\cite{qi2007experimental}.

CV-QKD systems with a locally generated LO at the receiver, also known as a real local oscillator (RLO) or local local oscillator (LLO) configuration, can eliminate side-channel attacks on the LO and offer a practical and simplified optical subsystem ~\cite{qi2015generating,huang2015high,kleis2017continuous, chin2021machine, laudenbach2019pilot}. However, LLO CV-QKD suffers from high excess noise caused by phase noise originating from the utilization of two independent lasers, limiting its transmission distance~\cite{marie2017self}. 

Two recent LLO implementations~\cite{pi2022sub, li2023continuous} have claimed success in their attempts to demonstrate CV-QKD over a long distance of 100 km. In an endeavor to tackle the phase noise issue, Yaodi \emph{et al}.~\cite{pi2022sub} employed an intricate system incorporating polarization multiplexing to separate a very strong pilot tone from the quantum signal, which makes their system suffer from some of the disadvantages of a TLO. Moreover, in Ref.~\cite{li2023continuous}, the authors used an arbitrary post-selection technique on data frames with low excess noise, leading to an underestimation of the excess noise and overestimation of the transmission range of the system. Finally, none of these works considered finite-size effects~\cite{scarani2009security}, a crucial aspect for practical applications. Consequently, practical and long-distance LLO CV-QKD is yet to be achieved.

Here, we report to our knowledge the longest-distance experimental demonstration of LLO CV-QKD that implements the entire QKD protocol and generates keys while taking finite-size effects into account. Specifically, we achieved a secret key rate of 25.4 kbits/s over 100 km of ultra-low-loss optical fiber with a total loss of 15.4 dB. This was made possible by controlling the excess noise using a machine learning (ML) framework for phase compensation~\cite{chin2021machine} and optimizing the modulation variance for information reconciliation. In our experiment, we consider multi-dimensional (MD) information reconciliation using a multi-edge-type low-density-parity-check (MET LPDC) error-correcting code~\cite{mani2021multiedge} with an efficiency of 92.5\%.

 \begin{figure*}[h]
\centering
\includegraphics[width=\linewidth]{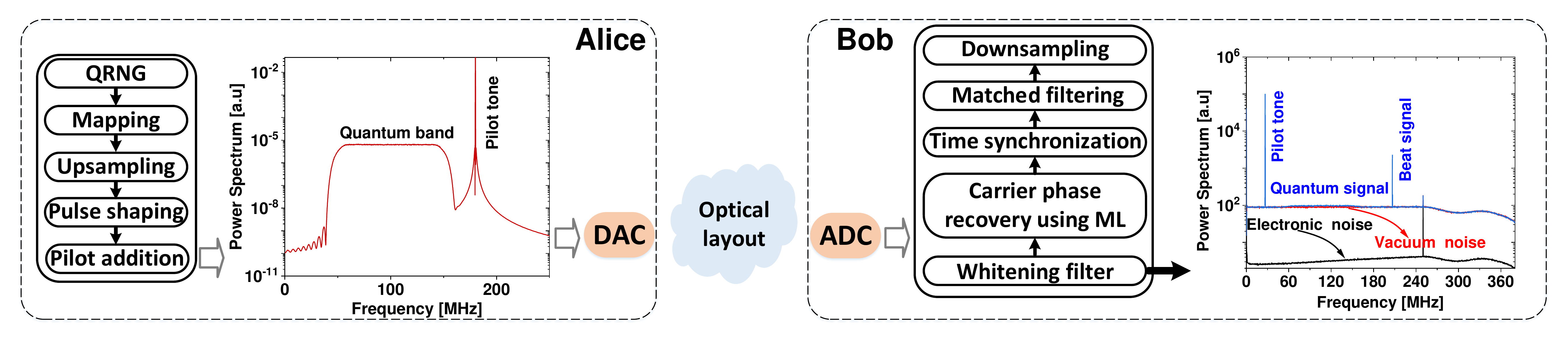}
\caption{\textbf{ Digital signal processing routines of the long-distance LLO CV-QKD system}. See the main text for the details.} 
\label{fig:2}
\end{figure*}
\section{Residual phase noise}
Excess noise in CV-QKD systems can arise from various sources, including quantization, modulation, relative intensity noise (RIN), Raman scattering, and residual phase noise. These noise sources are assumed to be statistically independent, and therefore the total excess noise can be expressed as the sum of individual contributions~\cite{laudenbach2018continuous}: 
\begin{equation}\label{eq:eq1}
\xi = \xi_{\text{RIN}}+\xi_{\text{mod}}+\xi_{\text{quant}}+\xi_{\text{Ram}}+\xi_{\text{RPN}} + \ldots \ 
\end{equation}
Among these noise sources, residual phase noise (RPN), defined as the variance of the difference between the actual phase of the quantum signal and the estimated phase of the received signal, is the main source of excess noise in LLO CV-QKD. In the Gaussian-modulated coherent state protocol, the excess noise due to the RPN at the receiver side can be expressed as~\cite{marie2017self},
\begin{equation}\label{eq:eq2}
\xi_{\text{RPN}} = 2TV_{\text{mod}}\left(1-e^{-\frac{{V_\text{RPN}}}{2}}\right),
\end{equation}
where $T$ is the transmittance, including the quantum channel and the detector efficiency, $V_{\text{mod}}$ represents the modulation variance, i.e.\ the variance of the coherent state ensemble, and $V_{\text{RPN}}$ denotes the variance of the RPN.

Following Eq.~\ref{eq:eq2} two options are available to reduce excess noise in LLO CV-QKD: operating the system at a low modulation variance or minimizing RPN. Although the former option is practical and straightforward to implement, it necessitates meticulous optimization of $V_{\text{mod}}$ due to the dependence of the secret key rate on the modulation variance. In particular, both the mutual information and the efficiency of information reconciliation are influenced by the modulation variance.  

Effective phase estimation is required to reduce RPN. Currently, the standard approach is to use pilot-aided techniques to estimate the relative phase between the free-running lasers of the transmitter and the receiver~\cite{qi2015generating,huang2015high,kleis2017continuous}. The quality of the estimated phase depends heavily on the signal-to-noise ratio (SNR) of pilot-aiding signals, implemented using single-frequency tones or training symbols, transmitted together with the quantum signal. However, these methods are limited by channel loss, which increases with distance, and the need for a low-power pilot to reduce cross-talk to the quantum signal. 
 
In contrast, machine learning (ML) has shown consistently excellent phase estimation performance across a wide range of pilot SNRs~\cite{chin2021machine}. This work combines ML-based phase estimation and modulation variance optimization to control excess noise to enable LLO CV-QKD over long distances.

\section{Experimental implementation}
\subsection{Optical layout}
Figure~\ref{fig:1} shows the optical layout of our long-distance LLO CVQKD system based on the Gaussian modulated coherent state protocol. At the sender, Alice, a continuous wave (CW) laser with a narrow line width of $\approx 100$ Hz and operating at a wavelength of 1550 nm was used as an optical carrier. The coherent states were prepared by modulating the CW laser using an in-phase and amplitude (IQ) modulator driven by a 16-bit digital-to-analog converter (DAC) with two channels operating at a sampling rate of 1 Gsample/s. The IQ modulator was operated in single sideband mode by controlling the direct current (DC) bias voltages using an automatic bias controller (ABC). A variable optical attenuator (VOA) was placed after the IQ modulator to adjust the modulation variance of the thermal state. A Faraday isolator was added at the sender output to avoid any back-reflections from the channel and Trojan-horse attacks. The signal was sent through a quantum channel made of a commercial ultra-low-loss fiber (TeraWave® SCUBA 150 Ocean Optical Fiber). The fiber attenuation is 0.146 dB/km at 1550 nm. The total loss in our 100 km fiber channel was 15.4 dB due to the mode field diameter difference between the SMF28 fiber pigtail and SCUBA 150. 
 
At the receiver, Bob, radio frequency (RF) heterodyne detection was used for the quantum state measurement. To accomplish this, another CW laser, free-running with respect to Alice's laser, was used as the LLO. The frequency difference between Alice's and Bob's lasers was $\approx 230$ MHz. The polarization of the quantum signal was then tuned to match the polarization of the LLO using a polarization controller. Next, the quantum signal and the LLO were combined on a balanced beamsplitter, followed by a home-made balanced detector with a bandwidth of $\approx 365$ MHz to detect the interference pattern. Finally, the detected signal was digitized using a 16-bit analog-to-digital converter (ADC) with a sampling rate of 1 Gsample/s and recorded for offline digital signal processing. The ADC and DAC were synchronized using a 10 MHz reference clock (CLK).

The measurement time was divided into frames, each containing $10^7$ ADC samples. Three measurements were taken autonomously: quantum signal measurement, vacuum noise measurement (Alice's laser off, Bob's laser on), and electronic noise measurement (Alice's laser off, Bob's laser off). The clearance of the vacuum noise over the electronic noise was $\approx15$ dB in the frequency band of the quantum signal. To calibrate the $V_\text{mod}$ of the thermal state, we performed back-to-back (B2B) measurements, in which Alice and Bob were connected through a short fiber patchcord, and the VOA was finely tuned to set different $V_\text{mod}$ values. %After the B2B measurement, $V_\text{mod}$ was set to be 8.41 shot-noise units (SNU), and the 100 km fiber channel was connected.

\subsection{Digital signal processing}
Figure ~\ref{fig:2} shows the offline digital signal processing (DSP) routine used for digital waveform generation and quantum symbols recovery at Alice's and Bob's stations, respectively. To produce an ensemble of coherent states, a sequence of random numbers with Gaussian distribution was generated by mapping the uniformly distributed output of a quantum number generator (QRNG) based on vacuum fluctuation~\cite{Gehring2021qrng}. These numbers form the complex amplitudes of the quantum symbol $\alpha_i=x_i+ip_i$ used for IQ modulation. These quantum symbols were drawn at a symbol rate of 100 MBaud, up-sampled to 1 Gsample/s and pulse-shaped using a root-raised cosine (RRC) filter with a roll-off factor of 0.2. For single-sideband modulation, the quantum signal was frequency shifted to 100 MHz. To this passband quantum signal, a pilot tone was multiplexed in frequency at 180 MHz for frequency and phase estimation at the receiver. The spectrum of the generated digital waveform is shown in the left part of Fig.~\ref{fig:2}. Finally, Alice uploaded her waveform into the DAC to obtain the corresponding electrical analog signals.

Extensive frame-to-frame DSP was deployed to reconstruct the quantum symbols at the receiver. After the digitization process using the ADC, a frequency domain equalizer (whitening filter) was applied to the quantum signal, vacuum noise, and electronic noise measurements. The filter coefficients were computed by taking the inverse of the receiver frequency response and averaging over 1000 frames. This step is essential to remove any auto-correlation and thus preserve the condition of independent and identically distributed quantum symbols. The whitened spectrum of the quantum signal, vacuum noise, and electronic noise is shown on the right side of Fig.~\ref{fig:2}. To estimate the phase and frequency difference between Alice's and Bob's lasers, a bandpass filter of 1 MHz was used to extract the desired pilot tone. The phase profile was extracted by computing a Hilbert transform of the filtered pilot. The frequency offset was then estimated using a linear fit. Using the estimated frequency offset, the pilot tone was baseband transformed and then used as an input signal to a ML framework based on an unscented Kalman filter (UKF) for phase estimation~\cite{chin2021machine}. After phase estimation, the quantum signal was shifted to baseband using the pilot frequency estimate and the known frequency offset between the quantum signal and pilot tone. The estimated phase from the UKF was used for correcting the phase of the quantum signal. The cross-correlation between reference transmitted samples and the receiver samples was used to compensate for the propagation delay of the fiber channel and different electronic components. Finally, the quantum symbols were recovered after matched RRC filtering and down-sampling.  

\subsection{Classical data processing}
\label{Ch: Classical data processing}
After DSP, Alice and Bob perform classical data processing, including information reconciliation, parameter estimation, and privacy amplification~\cite{jain2022practical}. For information reconciliation, we employed MD reconciliation based on MET-LDPC error-correcting codes with a rate of 0.05~\cite{mani2021multiedge}. While this code is designed to ideally operate at a fixed SNR, in the experiment, the SNR can vary, e.g., due to polarization fluctuations. As a result, we achieved a reconciliation efficiency of $\beta=$ 90.91\% and a frame error rate (FER) of 0\%.

%\begin{align}
%\nu(\mathbf{r,x}) =~0.05625~r_1x_1^{2}x_2^{20} + 0.04375~r_1x_1^{3}x_2^{25} +0.90~r_1x_3 \nonumber \\ 
%\varrho(\mathbf{x}) = 0.0265625x_1^{3} + 0.0234375x_1^{7}+0.48125x_2^{2}x_3^1+ 0.41875x_2^{3}x_3^1
%\end{align}

To enhance the reconciliation efficiency, we implemented a rate-adaptive reconciliation protocol that utilizes puncturing techniques to dynamically adjust the rate of the MET-LPDC code~\cite{mani2021multiedge,jain2022practical,hajomer2022modulation}. With this technique we achieved $\beta = 92.5\%$ and FER of 0.59. It is possible to achieve even higher efficiency of $ 93.1\%$ at the expense of increased FER of 0.80, but that actually reduces the length of the final secret key~\cite{hajomer2022modulation}.  % The code rate is defined as $R = {k}/{n}$, where \textit{n} is codeword length and \textit{k} denotes information bits. After puncturing, the final code rate can be changed adaptively according to the puncturing length $p$ as, $R_\text{punc}={k}/{(n-p)}~$. Thus, the efficiency of  MD reconciliation with a dimension dim = 8 becomes ${R_\text{punc}}/{C_{\text{AWGN}}(s)}$. By setting the puncturing length to 42000, corresponding to $R_\text{punc} = 0.0527$, we achieve $\beta = 93.7\%$ and FER of 92\%. However, achieving higher efficiency with a longer puncturing length is possible at the cost of high FER, which reduces the final secret key length. 

After error correction, parameter estimation was performed to evaluate the information advantage of the communicating parties over Eve by computing the Holevo bound. To accomplish this, we utilized all symbols, including the symbols of erroneous frames, i.e., the frames that Alice could not successfully decode. Finally, we applied privacy
amplification to generate the final key \cite{Tang2019}.   

\section{Results}
\begin{figure}
    \centering
    \includegraphics[width=.9\linewidth]{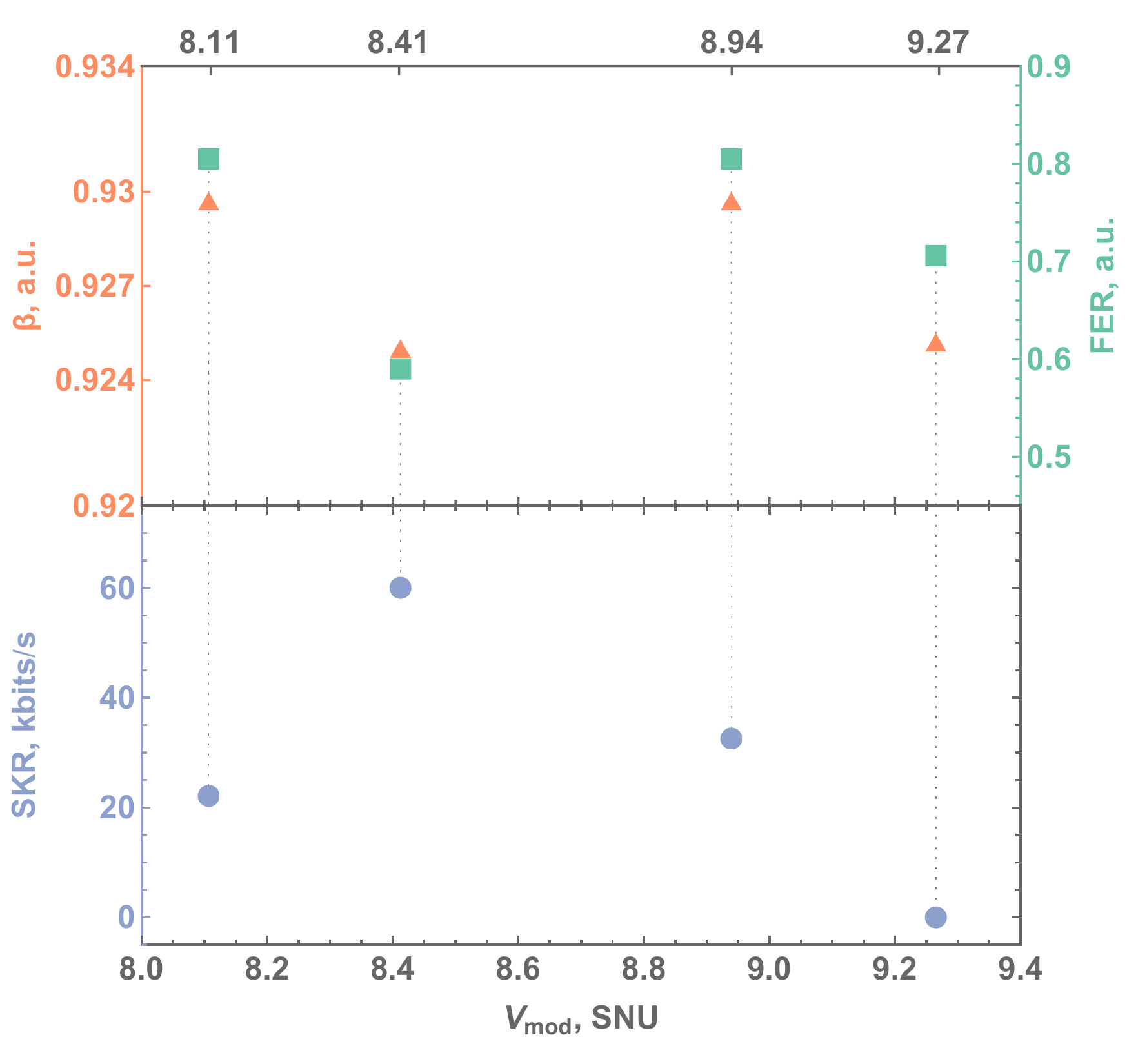}
    \caption{\textbf{Modulation variance optimization}. Experimentally obtained reconciliation efficiency $\beta$, FER and asymptotic SKR against the modulation variance $V_\text{mod}$.}
    \label{fig:4}
\end{figure}
As the first step, we optimized the modulation variance for the given MET-LDPC code with a rate of 0.05. This was done by performing information reconciliation on four sets of measurements, each consisting of $10^8$ symbols, which were taken at different modulation variances. Figure~\ref{fig:4} illustrates the figures of merit for information reconciliation ($\beta$ and FER) and the overall system performance (SKR in the asymptotic regime) as a function of $V_{mod}$. The highest efficiency of $\approx93\%$ with FER of 0.8 was obtained at $V_{mod}= 8.11$ SNU. However, this does not yield the best system performance in terms of SKR due to the high FER. Conversely, operating at the highest modulation variance of 9.27 SNU resulted in a lower FER of 0.7, but a null SKR due to the dominant effect of excess noise on system performance, as indicated by Eq.~\ref{eq:eq2}. Optimal system performance, striking a balance between $\beta$ (92.5\%) and FER (0.59) while maintaining low excess noise, was achieved by operating the system at $V_{mod}= 8.41$ SNU.

To evaluate the system performance, we used a security model with trusted devices~\cite{scarani2009security, jouguet2012analysis}, assuming that some noise and loss are inaccessible to Eve. Table~\ref{tab:1} summarizes the parameters used for secret key calculation. Alice generated an ensemble of $1\times10^9$ coherent states at a symbol rate of 100 MBaud, with modulation variance of 8.41 SNU, and transmitted them through a quantum channel with a mean untrusted transmittance of $\eta = 0.028$ and a mean excess noise of $\xi=0.212$ mSNU (at the channel output). The electronic/trusted noise of the detector and its efficiency/trusted transmittance had mean values of 62.72 mSNU and 0.68, respectively. For information reconciliation, Alice and Bob used $9.5\times10^8$ symbols, with some symbols discarded due to time synchronization. With puncturing, we achieved an efficiency of 92.5\%  and an FER of 0.59. %Nevertheless, improved FER performance can be anticipated when employing codes with optimal code rates at the operating SNR.

\begin{table}[t]
\centering
\caption{\textbf{Final experimental parameters}. $\tau$: Trusted efficiency, $\eta$: Untrusted efficiency, $t$: trusted detection noise, $\xi$: excess noise, FER: frame error rate, $\beta$: IR efficiency.}
\resizebox{0.98\hsize}{!}{
\begin{tabular}{cccc}
\hline
\bf Alice & \bf Bob&\bf channel &\bf IR\\
\hline
$B = 100$ MBaud &$\tau=0.68$ & $\eta=0.028$ & FER $=0.59$\\
$V_{mod} = 8.41$ SNU &$t=62.72$ mSNU &$\xi=0.212$ mSNU &$\beta = 92.5\%$ \\
\hline
\end{tabular}
}
  \label{tab:1}
\end{table}

\begin{table*}[h]
   \centering
\caption{\textbf{Comparison of long-distance CVQKD demonstrations}} \label{tab:table2}
%\resizebox{\hsize}{!}{
\begin{tabular}{|c|c|c|c|c|c|}
%         \hline  \multicolumn{3}{|c|}{\textbf{Entry}}  \\
         \hline \bf Ref.  &\bf Laser source & \bf LO & \bf Distance & \bf Modulation &\bf Security \\
       %  \hline  \cite{lodewyck2007quantum} & Pulsed & TLO & 25 km & Gaussian & Asymptotic& Applied\\ & \bf Information reconciliation 
       %  \hline  \cite{wang201525} & Pulsed & TLO & 50 km & Gaussian & Finite-size& Applied\\
         \hline  \cite{jouguet2013experimental} & Pulsed & TLO & 80 km & Gaussian& Finite-size\\
         \hline  \cite{huang2016long} & Pulsed & TLO & 100 km & Gaussian & Finite-size\\
         \hline  \cite{zhang2020long} & Pulsed & TLO & 202.18 km & Gaussian & Finite-size\\
         %\hline  \cite{qi2015generating,huang2015high} & Pulsed & LLO & 25 km & Gaussian& Finite-size& Not Applied\\
        % \hline  \cite{laudenbach2019pilot} & Pulsed & LLO & 40 km &  Discrete & Asymptotic & Not Applied \\
        % \hline  \cite{kleis2017continuous} & CW & LLO & 40 km & Discrete  & Asymptotic & Not Applied\\
        % \hline  \cite{chin2021machine} & CW & LLO & 20 km & Gaussian & Asymptotic & Not Applied\\
         % \hline  \cite{jain2022practical} & CW & LLO & 20 km & Gaussian & Composable & Applied\\
        % \hline  \cite{hajomer2022modulation} & CW & LLO & 20 km & Gaussian & Composable & Applied\\
         \hline  \cite{hajomer2022continuous} & CW & LLO & 60 km & Gaussian & Asymptotic\\
          \hline  \cite{pi2022sub} & CW & LLO & 100 km & Gaussian & Asymptotic\\
         \hline  \cite{li2023continuous} & CW & LLO & 100 km & Gaussian & Asymptotic \\
         \hline  Current work  & CW & LLO & 100 km & Gaussian& Finite-size \\
         
         \hline
\end{tabular}
%}
\end{table*}

 \begin{figure}[t]
\centering

\includegraphics[width=\linewidth]{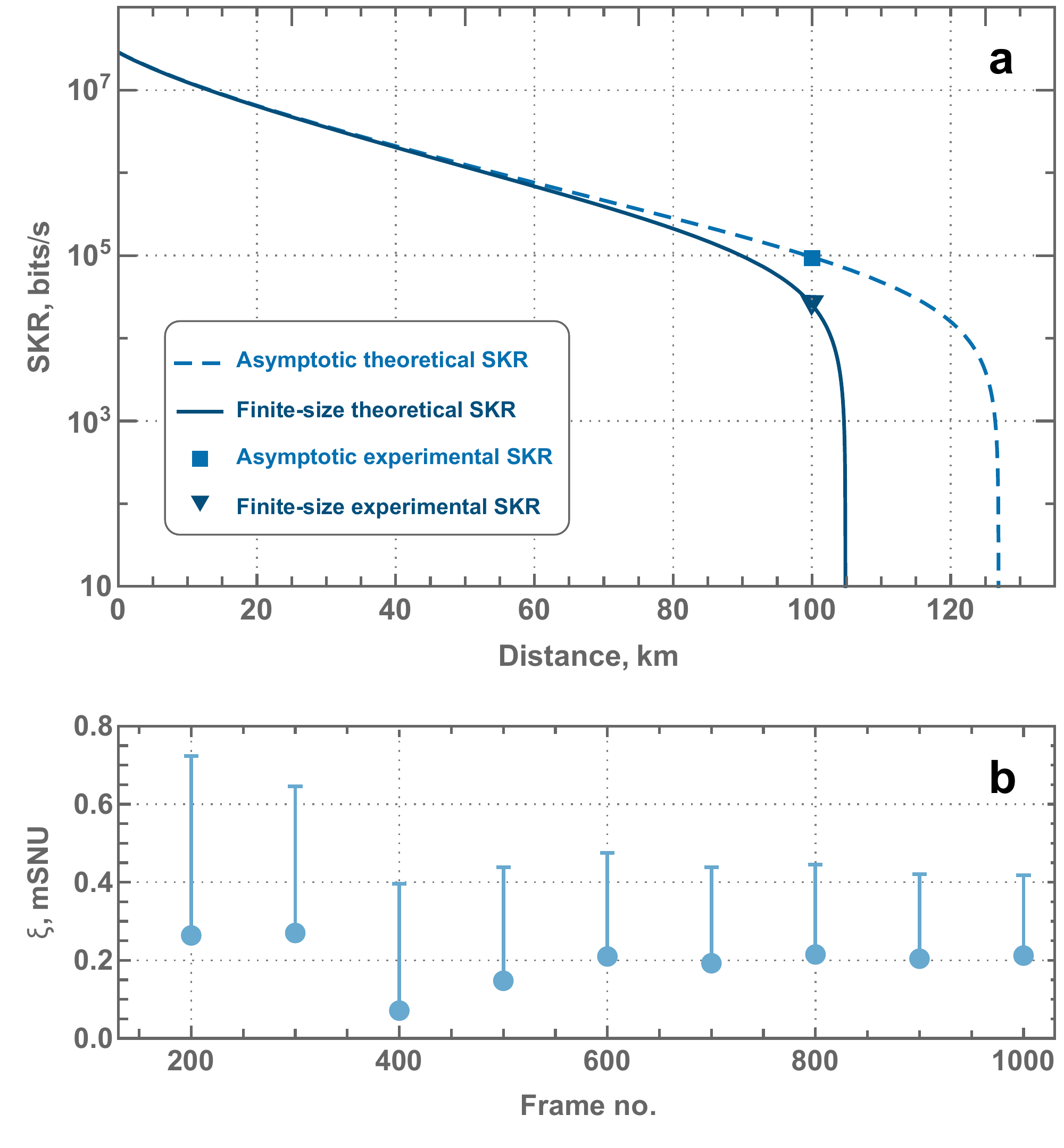}
\caption{\textbf{Performance of long-distance CV-QKD}. (a)  The secret key rate (SKR) versus fiber channel length with attenuation factor 0.146 dB/km (taking into account additional coupling loss of $82\%$) in asymptotic (dashed) and finite-size (solid) regimes. Points correspond to experimentally achieved results. (b) Cumulative excess noise as a function of the number of acquired frames.} 
%From top to bottom $\beta=96\%,\,95\%$ (dashed) and $93.7$\% (solid).
\label{fig:3}
\end{figure}

The key is deemed to be secure provided the positivity of the accessible information difference \cite{devetak2005distillation}: 
\begin{equation}
\label{eq:key}
SKR(\eta_\text{low},\xi_\text{up})=B\times(1-\text{FER})\left(\beta I_{AB} - \chi_E - \Delta(n)\right),
\end{equation}
where the mutual information between trusted parties $I_{AB}$ and the upper bound on the information attainable by an eavesdropper $\chi_E$ are evaluated using $\eta_\text{low}$ and $\xi_\text{up}$ worst-case estimates of channel loss and excess noise respectively, taken with error probability (i.e., the probability that true value falls outside the Gaussian confidence interval) of  $\delta_{fail}=10^{-10}$ \cite{ruppert2014}. The correction term $\Delta(n)$ is related to the security of the privacy amplification procedure \cite{leverrier2010finite} on a key block of length $n$ that is used to establish the key. In the asymptotic regime, the correction term $\Delta(n)$ is neglected, and true channel parameters $\eta$ and $\xi$ are presumed to be known and equal to estimated values. \par

The secret key rates for both numerical simulation and experimental results are depicted in Fig.~\ref{fig:3} (a). The square and triangle correspond to our experimental results in asymptotic and finite-size regimes, respectively. A secure key generation rate exceeding $25$~kbits/s was achieved in both regimes. For given experimental parameters, one can theoretically predict a secure distance beyond 100 km. 

Figure \ref{fig:3} (b) shows the evolution of the excess noise with the frames accumulated and processed, with the respective upper bound of the confidence interval indicated by the dash. The protocol's performance could be further improved by accumulating more data, consequently leading to tighter confidence in estimated channel noise. %for $\beta=93.1\%$ and $\beta=92.5\%$ (with corresponding FER, see Table \ref{tab:1} for details), respectively. %The code efficiencies $\beta$ and FERs are given in Table \ref{tab:1}. , our setup can achieve 18.67 kbps in the asymptotic regime. In the finite-size regime, the same code could be used to achieve 1.546 kbps albeit with a significantly larger magnitude of error $\delta_\text{fail}=10^{-2}$. 
%Higher efficiency $\beta$ translates to higher key generation speed and asymptotically our setup can achieve 60~kbits/s and 95.9~kbits/s, respectively. After taking into account finite-size effects, our codes allow us to reach approximately the same secure key generation rate of $25.6$~kbits/s because of different FERs. Thus, the trade-off between error correction efficiency and the achievable FERS is necessary to be considered to optimize the final secret key length. However, 
%a more efficient code with $\beta=96\%$ \cite{zhang2020long,jouguet2011long, milicevic2018quasi} with the same rate $B$ (and FER=0.80) would allow an SKR up to 89.4~kbits/s. \par
%To account for finite-size effects, we considered a Gaussian confidence interval with an estimation failure probability (i.e. ) of $\delta_\text{fail}=10^{-9}$ \cite{ruppert2014}. After considering the worst-case estimator of the untrusted noise and the channel transmittance, we achieve a positive key rate of 40kbits/s and 2kbits/s for asymptotic and finite-size with a failure probability of $10^{-6}$ at a distance exceeding 100 km. Fig.2 (b) depicts the experimentally measured untrusted noise at the detector output for 100 frames each with a $10^6$ symbol and the corresponding worst-case estimator. 
\section{DISCUSSION}
 Long-distance transmission is a key requirement for large-scale deployment and integration of QKD in existing telecom networks. CV-QKD lends itself naturally to this integration. However, the secure and practical system configuration (LLO CV-QKD) faces limitations in transmission range due to the phase noise of lasers. In this work, we demonstrated long-distance LLO CV-QKD over a 100 km fiber channel, while accounting for finite-size effects. This record-setting experiment was made possible by using machine learning for phase noise compensation and optimizing the modulation for information reconciliation and excess noise simultaneously.  

Table~\ref{tab:table2} compares key aspects of long-distance CV-QKD experiments performed over the past ten years. To achieve secure key generation beyond 70 km, previous demonstrations have used the TLO configuration and pulse carving, which introduce vulnerabilities and requires an additional amplitude modulator, respectively. Thus far, the maximum distance of the LLO CVQKD experiment with actual key generation was 60 km. The recent demonstrations of 100 km  LLO CVQKD have not considered the finite-size regime, a crucial aspect for practical applications. The authors of Ref.~\cite{li2023continuous} used an unjustifiable post-selection technique on the frames with low excess noise, equivalent to underestimating the actual excess noise and overestimating the distance. Moreover, in Ref.~\cite{pi2022sub}, a DAC with only 8-bit resolution was used, indicating a poor approximation of the continuous Gaussian modulation. Furthermore, this work highlights the necessity of intricate system implementation, encompassing polarization multiplexing and an additional balanced detector, to effectively segregate the strong pilot tone from the quantum signal. Our work demonstrates the actual key generation in both asymptotic and finite-size regimes considering collective attacks. This achievement closes the gap between LLO- and TLO-CVQKD systems’ performance while maintaining a high level of security and lowering the implementation complexity.

Nonetheless, there is significant room for improvement in the current implementation. The overall system performance can be improved using MET-LDPC with a more suitable code rate of 0.03, allowing the system to operate at the optimal modulation variance of $\approx3.5$ SNU for $\beta = 92 \%$. %This can also extend distance further. 
Additionally, to achieve composable security, a higher symbol rate is necessary to collect a large number of symbols. This can be achieved by increasing the system's bandwidth, for instance, by using high-speed DACs and ADCs combined with a broadband balanced detector~\cite{bruynsteen2021integrated}. 

In summary, this experiment has the potential to pave the way for realizing CV quantum networks, such as quantum passive optical networks, where high loss tolerance and LLO are essential ingredients. We believe that this will ultimately be a key enabler for the large-scale deployment of secure quantum communication.

\begin{backmatter}
\vspace{0.5cm}
\bmsection{Data availability} Data underlying the results presented in this paper are available from the authors upon reasonable request.

\smallskip

\bmsection{Acknowledgments} We thank OFS optics for providing the SCUBA150 fiber for this experiment. All authors acknowledge support from the Innovation Fund Denmark (CryptQ, grant agreement no. 0175-00018A) and from the Danish National Research Foundation, Center for Macroscopic Quantum States (bigQ, DNRF142). This project was funded within the QuantERA II Programme (project CVSTAR) that has received funding from the European Union’s Horizon 2020 research and innovation program under Grant Agreement No 101017733. AH and TG acknowledge funding from the Carlsberg Foundation, project CF21-0466. ID acknowledges support from the project 22-28254O of the Czech Science Foundation.

\smallskip

\bmsection{Competing interests} The authors declare no competing interests.

\smallskip

\bmsection{Author contributions statement} A.A.E.H. performed the experiment and the overall data processing and analysis, with inputs from N.J. and H.-M.C. in the implementation of DSP routines. I.D. performed the theoretical analysis. N.J. performed error correction. H.-M.C. implemented the machine learning framework. T.G. contributed to all parts of the work. A.A.E.H. and T.G. wrote the manuscript. A.A.E.H. and T.G. conceived the experiment, and U.L.A. and T.G. supervised the project. All authors were involved in discussions and interpretations of the results.
\smallskip

\end{backmatter}

\bigskip

\bibliography{sample}

\bibliographyfullrefs{sample}

\ifthenelse{\equal{\journalref}{aop}}{%
\section*{Author Biographies}
\begingroup
\setlength\intextsep{0pt}
\begin{minipage}[t][6.3cm][t]{1.0\textwidth} % Adjust height [6.3cm] as required for separation of bio photos.
  \begin{wrapfigure}{L}{0.25\textwidth}
    \includegraphics[width=0.25\textwidth]{john_smith.eps}
  \end{wrapfigure}
  \noindent
  {\bfseries John Smith} received his BSc (Mathematics) in 2000 from The University of Maryland. His research interests include lasers and optics.
\end{minipage}
\begin{minipage}{1.0\textwidth}
  \begin{wrapfigure}{L}{0.25\textwidth}
    \includegraphics[width=0.25\textwidth]{alice_smith.eps}
  \end{wrapfigure}
  \noindent
  {\bfseries Alice Smith} also received her BSc (Mathematics) in 2000 from The University of Maryland. Her research interests also include lasers and optics.
\end{minipage}
\endgroup
}{}

\end{document}